\documentclass[11pt,twoside]{article}
\usepackage{asp2004}
\usepackage{epsf}
\usepackage{psfig}
\usepackage{lscape}
\begin{document}

\title{The Composition of the Sagittarius Dwarf Spheroidal Galaxy, and Implications 
          for Nucleosynthesis and Chemical Evolution   }

\author{Andrew McWilliam}
\affil{The Observatories of the Carnegie Institution of Washington, 813 Santa Barbara Street,
        Pasadena, CA 91101; andy@ociw.edu}

\author{Tammy A. Smecker-Hane}
\affil{Department of Physics and Astronomy, University of California, 4129 Frederick Reines Hall, 
       Irvine, CA 92697; tsmecker@uci.edu}

\begin{abstract}

We outline the results of a study of the chemical composition of 14 stars in the 
Sagittarius dwarf spheroidal galaxy (Sgr dSph).  For the Sgr dSph stars with [Fe/H]$\geq$$-$1 
the abundances are highly unusual, showing a striking enhancement in heavy s-process elements,
increasing with [Fe/H], deficiencies of the $\alpha$-elements (O, Si, Ca, and Ti), deficiencies of
Al and Na, and deficiencies of the iron-peak elements Mn and Cu.  Our abundances suggest that the
composition of the metal-rich Sgr dSph stars is dominated by the ejecta of an old, metal-poor 
population, including products of AGB stars and type~Ia supernovae (SN).

We suggest two scenarios to explain the observations: First, is chemical enrichment over long timescales
in a galaxy which has experienced significant mass-loss during its evolution.  The second possibility
is that we are seeing the products of chemical enrichment from a system which experienced 
a large burst of star-formation, followed by a quiescent period of may Gyr.  Both of these scenarios
can lead to the situation where newly synthesized material from AGB stars and type~Ia SN
overwhelms nucleosynthesis products from a minor population of young, metal-rich, stars.  It is
likely that both mechanisms operate in the Sgr dSph.

Since these conditions should be generally applicable to low-mass systems,
we expect to find similar abundance patterns in other dwarf galaxies.  This
is supported by the chemical composition of stars in the Galactic globular cluster $\omega$~Cen and
in the Fornax dwarf galaxy, suggesting that both shared a history similar to the Sagittarius dwarf
spheroidal galaxy.

\end{abstract}

\section{Introduction}

In this paper we discuss results from our abundance study
of Sgr dSph red giant stars.  Abundance results for the $\alpha$-element
group excluding oxygen (Ca, Si, Ti), Fe, Na, Al, and the neutron-capture
elements Y, La and Eu are taken from Smecker-Hane \& McWilliam (2002, henceforth SM02; 2005
in preparation).  The abundances for the iron-peak element Mn were taken
from McWilliam, Rich, \& Smecker-Hane (2003), whilst the results for Cu are taken from
McWilliam \& Smecker-Hane (2004).  The oxygen results were measured for
this paper, but will be discussed in more detail in a later publication.

\section{Observations and Analysis}

\begin{figure}[!ht]
\plotone{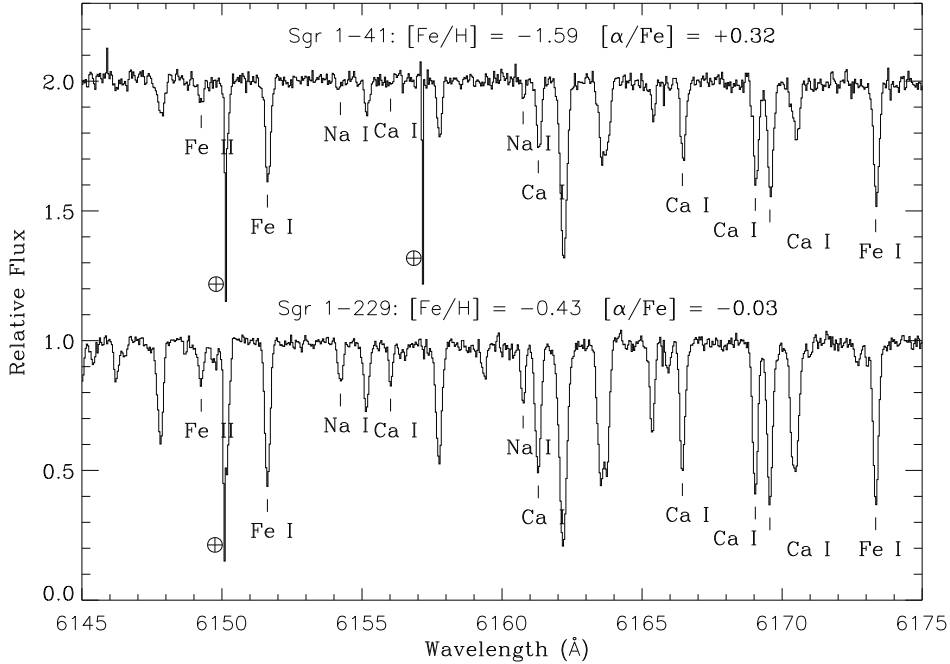}
\caption{Spectra of a metal--poor (top) and metal--rich (bottom) Sgr dSph
star. Spectra are shown shifted to rest wavelengths, measured lines
are labeled, and the derived metallicities are shown. Bad columns on
the CCD are designated with the $\oplus$ symbol.}
\end{figure}

Details of the spectral observations of 14 Sgr dSph red giant stars are described in
SM02 and Smecker-Hane \& McWilliam (2005, in preparation).  Briefly, we obtained
Keck~I\footnote{The W.M. Keck Observatory is operated as a scientific partnership among the
California Institute of Technology, the University of California, and the
National Aeronautics and Space Administration.  The observatory was made possible
by the generous financial support of the W.M. Keck Foundation.  We extend special
thanks to the people of Hawaiian ancestry on whose sacred mountain we were privileged to
be guests.}, HIRES (Vogt et al. 1994), spectra with resolving power R$\sim$34,000
and S/N$\sim$50 per extracted pixel.  Extractions were performed using the IRAF suite of routines.
Examples of typical spectra are shown in Figure~1.

Equivalent widths (EW) of spectral lines were measured using the semi-automated
routine GETJOB (McWilliam et al. 1995).  Abundances were calculated from the EWs using the
spectrum synthesis program MOOG (Sneden 1973).  For the Cu~I line at 5782\AA , and other
lines affected by hyperfine splitting, it was necessary to use the MOOG {\em blends} option
to properly synthesize the feature EW.  Model atmospheres were linearly interpolated
from the grid of Kurucz (1993) 64-layer atmospheres with convective overshoot.

\section{Abundance Results}

Figure~2 shows our abundance results for the average of three alpha-elements
(Ca, Si and Ti) with [Fe/H], compared to the locus for the solar neighborhood,
found by Edvardsson et al. (1993); it is clear that [$\alpha$/Fe] is
deficient in the Sgr dSph, relative to the local trend.   The alpha-element deficiency in
Sgr dSph corresponds to a shift of the trend by $\sim$0.4 dex in [Fe/H].  Figure~2 also
shows that the Sgr dSph stars below [Fe/H]$\sim$$-$1 have alpha enhancements of
$\sim$0.3 dex, typical of the Galactic halo.  It is encouraging that the composition
of two of the Sgr dSph globular clusters (M54 and Ter~7) are consistent with our individual
field stars.  The two stars studied by Bonifacio et al. (2000) appear to have slightly
lower $\alpha$/Fe ratios than our points.

It is interesting that the [Fe/H] values of our Sgr dSph stars ranges from $-$1.6
to $+$0.0 dex, which is quite large considering that Sgr dSph is a dwarf galaxy.  We note
that our target selection was biased, based on approximate metallicities from 
Ibata (private communication), to cover a large range of metallicity
and to focus more on stars within 0.5 dex of the solar value, in order to
compare our abundance results to the trend of $\alpha$/Fe in the solar neighborhood.

\begin{figure}[!ht]
\plotone{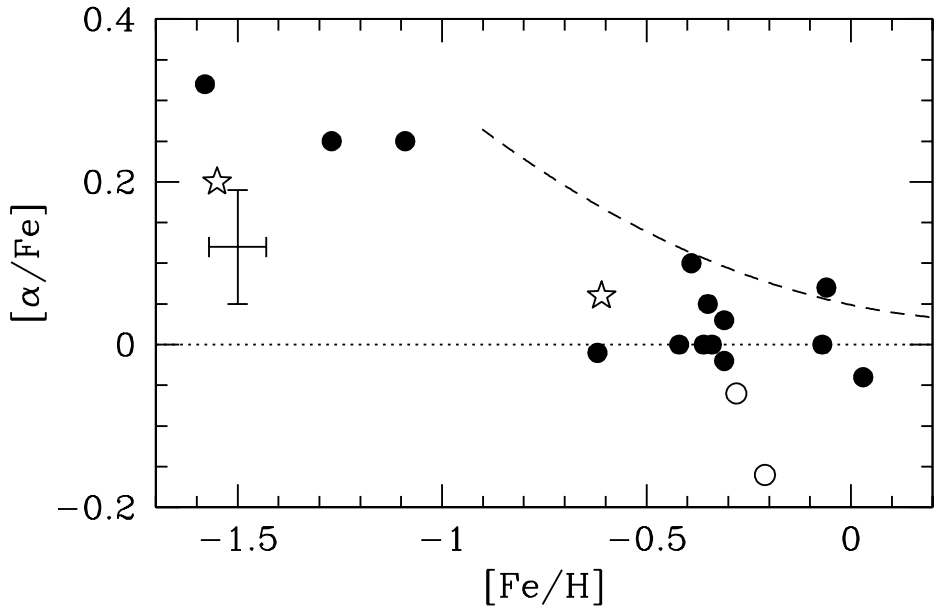}
\caption{Alpha-element abundances for red giants in the Sgr dSph (filled circles),
where [$\alpha$/Fe] is the average of [Si/Fe], [Ca/Fe], and [Ti/Fe].
A typical error bar shown on the lower left. The average abundance of
red giants in M54 and Ter~7, globular clusters in the Sgr dSph,
from Brown, Wallerstein \& Gonzalez (1999) and Tautvaisiene et al. (2004) respectively,
are shown as open star symbols.  Sgr dSph abundances from Bonifacio et al.~(2000) 
are shown as open circles.  The dashed line represents the mean trend in [$\alpha$/Fe] for
stars in the solar neighborhood from Edvardsson et al.~(1993). }

\end{figure}

The trend of oxygen in the Sgr dSph stars, shown in Figure~3, is consistent with the 
$\alpha$-element trend of Figure~2.  For the oxygen abundances in Figure~3 we have
used the the same $gf$ values and Ni~I blend for the [O~I] 6300\AA\ line as utilized
by Allende Prieto, Lambert, \& Asplund (2001).  We computed the solar oxygen abundance using 
the Kurucz grid of 1D models with overshoot, rather than the 3D hydrodynamical model
with non-LTE employed by Allende Prieto et al. (2001); however, our result for the
solar [O~I] 6300\AA\ line is very close to their value (they found 
log~$\epsilon$(O)=8.69$\pm$0.05 dex, whereas we calculated  8.73).  
We note that for the measurement of the solar [O~I] line we used the same continuum
regions as indicated by Allende Prieto et al. (2001).

In Figure~3 we compare the [O/Fe] values found here for the Sgr dSph stars with
Allende Prieto et al. (2004) results for the solar neighborhood stars within 14.5 pc.
The Sgr dSph stars show a similar slope of [O/Fe] with [Fe/H] for the metal-rich stars, 
but shifted to lower [Fe/H] than in the solar neighborhood, with a trend that extends
almost 0.2 dex below the solar [O/Fe] value.  This behavior is roughly similar to the
the trend of the other $\alpha$-elements above [Fe/H]$\sim$$-$0.6.  An alternative
description is that the [O/Fe] trend
appears to be shifted by approximately 0.3 dex in [Fe/H] relative to Allende Prieto
et al. (2004).

We note that two of our three Sgr dSph stars with [Fe/H] $\leq$$-$1 show halo-like 
oxygen enhancements.  The metal-poor Sgr dSph star, I-73, does not appear on Figure~3
because its very deficient oxygen abundance ([O/Fe]=$-$0.91 dex), puts it outside our
chosen plot boundary.  

The low [$\alpha$/Fe] and [O/Fe] ratios observed in Sgr dSph can be understood in a paradigm,
suggested by Tinsley (1979; but see also Wheeler, Sneden \& Truran 1989, and McWilliam 1997),
to explain the trend of decreasing [O/Fe] with increasing metallicity
from $-$1$\leq$[Fe/H]$\leq$0.  In this scenario the low metallicity, halo, stars have 
high [$\alpha$/Fe], characteristic of ejecta from core-collapse type~II SN, because the
type~II SN progenitors are short-lived and are first to eject their material back into
the ISM.  Type~Ia SN add to the chemical composition on long time scales, of order one to
a few Gyr.  Since type~Ia produce low $\alpha$/Fe material, the composition of stars formed
over long time scales show lower [$\alpha$/Fe] ratios.  In this paradigm the [Fe/H] value
at which the [$\alpha$/Fe] begins to decline from the halo ratio depends upon the star
formation rate: Slowly evolving systems are expected to show the decline in [$\alpha$/Fe]
at lower [Fe/H], because they do not reach as high [Fe/H] before the time when type~Ia SN
begin to contribute significantly to the ISM composition.  Thus, our observations of
[$\alpha$/Fe] and [O/Fe] indicate that the Sgr dSph had a lower star formation rate than
the solar neighborhood experienced, and that significant contribution from type~Ia SN had 
occurred over a long timescale.  These results are expected for low-mass galaxies (e.g. see 
Matteucci \& Brocato 1990).

\begin{figure}[!ht]
\plotone{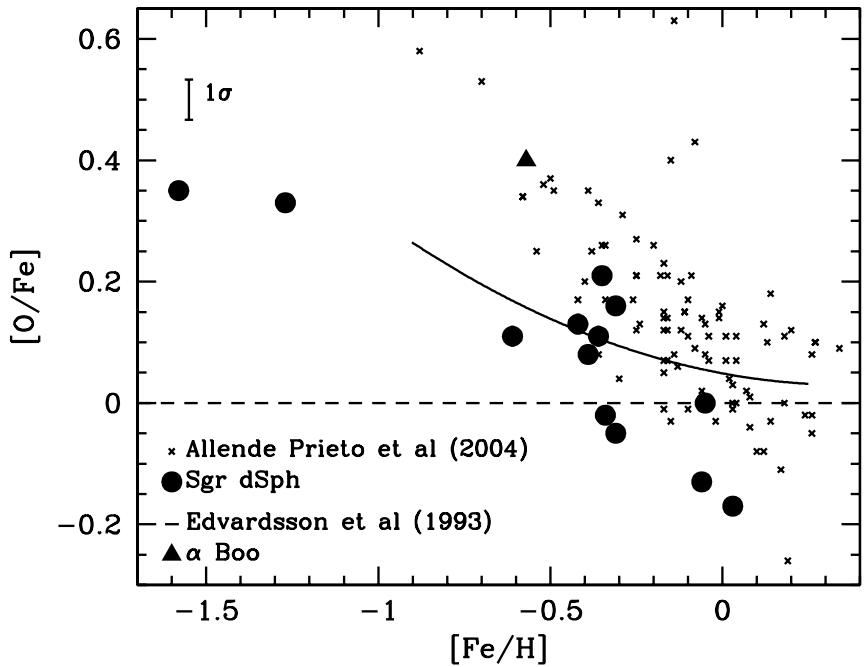}
\caption{Oxygen abundances in Sgr dSph red giant stars (filled circles), compared with
the solar neighborhood results of Allende Prieto et al. (2004, small crosses), and
the $\alpha$-element trend of Edvardsson et al. (1993, solid line).  The location
of $\alpha$~Boo (filled triangle) was computed by McWilliam (2004 in progress).
%using the Hinkle et al. (2000) atlas.
}
\end{figure}

Figure~4 shows the trends of [Al/Fe] and [Na/Fe] in the Sgr dSph, compared to solar neighborhood
results from Chen et al. (2000); the plot is taken from SM02.  Both Al and Na are
deficient in the Sgr dSph stars compared to the solar neighborhood: Sgr dSph stars with
[Fe/H]$\geq$ $-$0.6 dex are typically deficient in [Al/Fe] by $\sim$0.3--0.4 dex, and deficient
in [Na/Fe] by $\sim$0.4 dex.

Among the Sgr dSph stars with [Fe/H]$\leq$$-$1 dex two points are missing from Figure~4:  For 
star I-87 the 6696 and 6698\AA\ Al~I lines were not detected, whilst star I-73 has such a large
Al enhancement, at [Al/Fe]=$+$1.12  dex, that the scale of the figure would be drastically 
altered in order to include this data point, and would mask the contrast between the general
[Al/Fe] level in the Sgr dSph and solar neighborhood stars.

Because star I-73 shows a large enhancement of aluminum and a large deficiency of oxygen,
together with a mild sodium enhancement, we suspect that its atmospheric
composition is contaminated by proton-burning products, often seen
in Galactic globular clusters (e.g. Kraft el al. 1997, Sneden et al. 2004).  Our
iron abundance for I-73,  at [Fe/H]=$-$1.09,  differs significantly
from the M54 [Fe/H] value, of $-$1.55, found by Brown et al. (1999); thus, I-73 is probably
not a member of M54, and is the only known field star showing the O-Al anti-correlation.
Star I-73 might be a member of M54 if either the Brown et al. (1999) metallicity scale
differs from ours, or if M54 possesses a metallicity spread, similar to the
Galactic globular cluster $\omega$~Cen, with star I-73 simply a high-metallicity M54 star.

The main bulk of Na and Al production occurs during hydrostatic burning phases
in massive stars, which end as type~II SN (e.g. Timmes, Woosley, \& Weaver 1995;
Arnett \& Thielemann 1985; Woosley \& Weaver 1995).  The slight increase in [Na/Fe]
and [Al/Fe] seen in disk stars from [Fe/H]=0 to $-$1 (Reddy et al. 2003, Edvardsson et al. 1993),
the [Al/Fe] enhancement in thick disk stars (Reddy et al. 2003), and the [Al/Fe] enhancement
in Galactic bulge stars (McWilliam \& Rich 2003), are all consistent with these two elements 
being mildly $\alpha$-like, mostly made by type~II SN.

$^{23}$Na is produced as a primary product of the carbon-burning shell, although the final
yields are metallicity-dependent (Woosley \& Weaver 1995).  Observationally, over the full
range of [Fe/H] observed in the Galaxy [Na/Fe] roughly scales with [Fe/H] (e.g.  Pilachowski,
Sneden \& Kraft 1996; McWilliam 1997), conflicting with the Timmes et al. (1995) predictions
of metallicity-dependence.  

The yield of $^{27}$Al in the carbon-burning shell depends on the metallicity of the star,
whereas, in the neon-burning shell weak interactions transform the neutron excess, so that there
the $^{27}$Al yield is not so sensitive to metallicity (e.g. Woosley \& Weaver 1995).
Observations of Galactic halo field stars  (e.g. Gratton \& Sneden 1988; Shetrone 1996)
show that from [Fe/H]=$-$1 to $-$3 the halo field stars have a strongly metallicity-dependent
[Al/Fe] ratios, even more sensitive to [Fe/H] than the theoretical expectations of
Timmes et al. (1995).

We interpret the low [Na/Fe] and [Al/Fe] ratios as most likely the result of a
diminished role of type~II SN nucleosynthesis products in the composition
of the metal-rich Sgr dSph stars, similar to the conclusion reached from the
$\alpha$-elements; although the Na and Al deficiencies are larger than for
the $\alpha$-elements.  However, the possibility of metallicity-dependent Na and Al
yields complicates interpretation of the abundances.

\begin{figure}[!ht]
\plotone{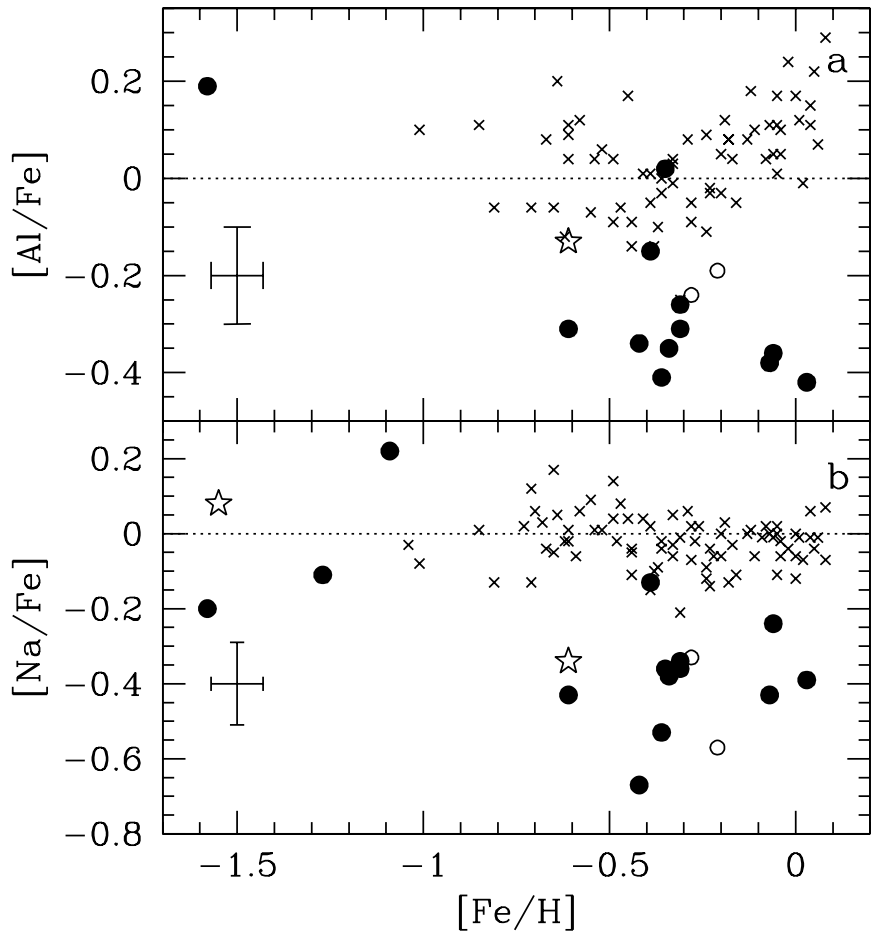}
\caption{(From SM02)\quad{\bf a: }A plot of [Al/Fe] versus [Fe/H].
{\bf b: }A plot of [Na/Fe] versus [Fe/H].
Symbols are the same as in Figure~2. Crosses
represent abundances from Chen et al.~(2000) for
solar neighborhood F stars.}
\end{figure}

In Figure~5 (taken from SM02) we show one of the most striking
abundance trends present in the
Sgr dSph stars: the strong enhancement of heavy neutron-capture element abundances with
increasing [Fe/H]; note that the most metal-rich stars have enhancements up to $\sim$$+$1
dex above the solar value.  It is significant that the largest neutron-capture enhancements, seen
in the most metal-rich Sgr dSph stars, show a scatter larger than the estimated abundance
uncertainties.  The, normally modest, La~II lines appear strong
in most of our spectra, which consequently required careful abundance analysis, including the
treatment of hyperfine splitting ($hfs$).   It would have been useful to 
measure barium abundances from the Ba~II lines; but all those lines were very strong, 
on the flat portion of the curve of growth, resulting in highly uncertain abundances.

%Open squares in Figure~5 show the solar neighborhood results from a sample of stars
%studied by Sneden and Gratton (***get refs! ****???); more recent results, for example
%by Simmerer et al. (2004?) are now published, but were not available at the time this
%plot was made (by SM02).

\begin{figure}[!ht]
\plotone{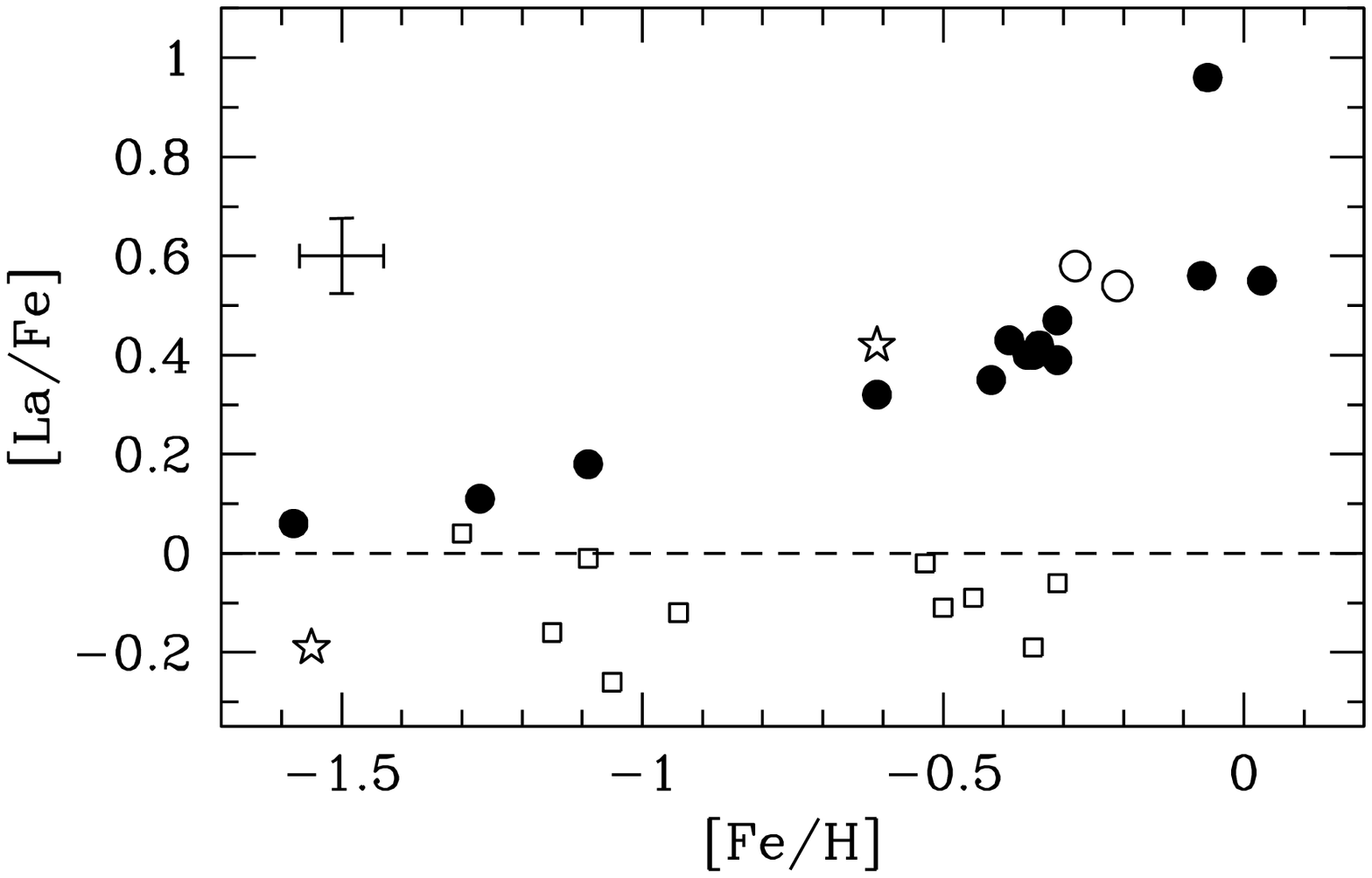}
\caption{ (From SM02)\quad A plot of [La/Fe] versus [Fe/H].
Symbols are the same as in Figure~2.
Open squares represent chemical abundances
of Galactic stars from Gratton \& Sneden (1994). }
\end{figure}

Figure~6 shows our abundance results for [Eu/Fe], based on the Eu~II line at 6645\AA ; the
analysis included appropriate $hfs$ treatment.  The figure also indicates the results for the 
solar neighborhood [Eu/Fe] ratios from a variety of papers, which demonstrate that the Eu 
overabundances in metal-poor stars follow a trend much like that of the $\alpha$-elements.  
It is clear that the [Eu/Fe] ratios in
the Sgr dSph are $\sim$0.5 dex higher than in the solar neighborhood, at solar metallicity.
Figure~7 shows that the ratio of [La/Eu] trends toward the s-process value with increasing
[Fe/H], whilst at low [Fe/H] the [La/Fe] ratios resemble more the r-process dominated
composition of Galactic halo stars.  Together, these two figures show that [Eu/Fe] can be enhanced
even in s-process dominated material, and that, therefore, it is dangerous to blame the
r-process as responsible for heavy-element nucleosynthesis based on [Eu/Fe] alone; instead, it
is necessary to ratio the abundances of two heavy elements which have different sensitivities
to the neutron-capture timescale.

\begin{figure}[!ht]
\plotone{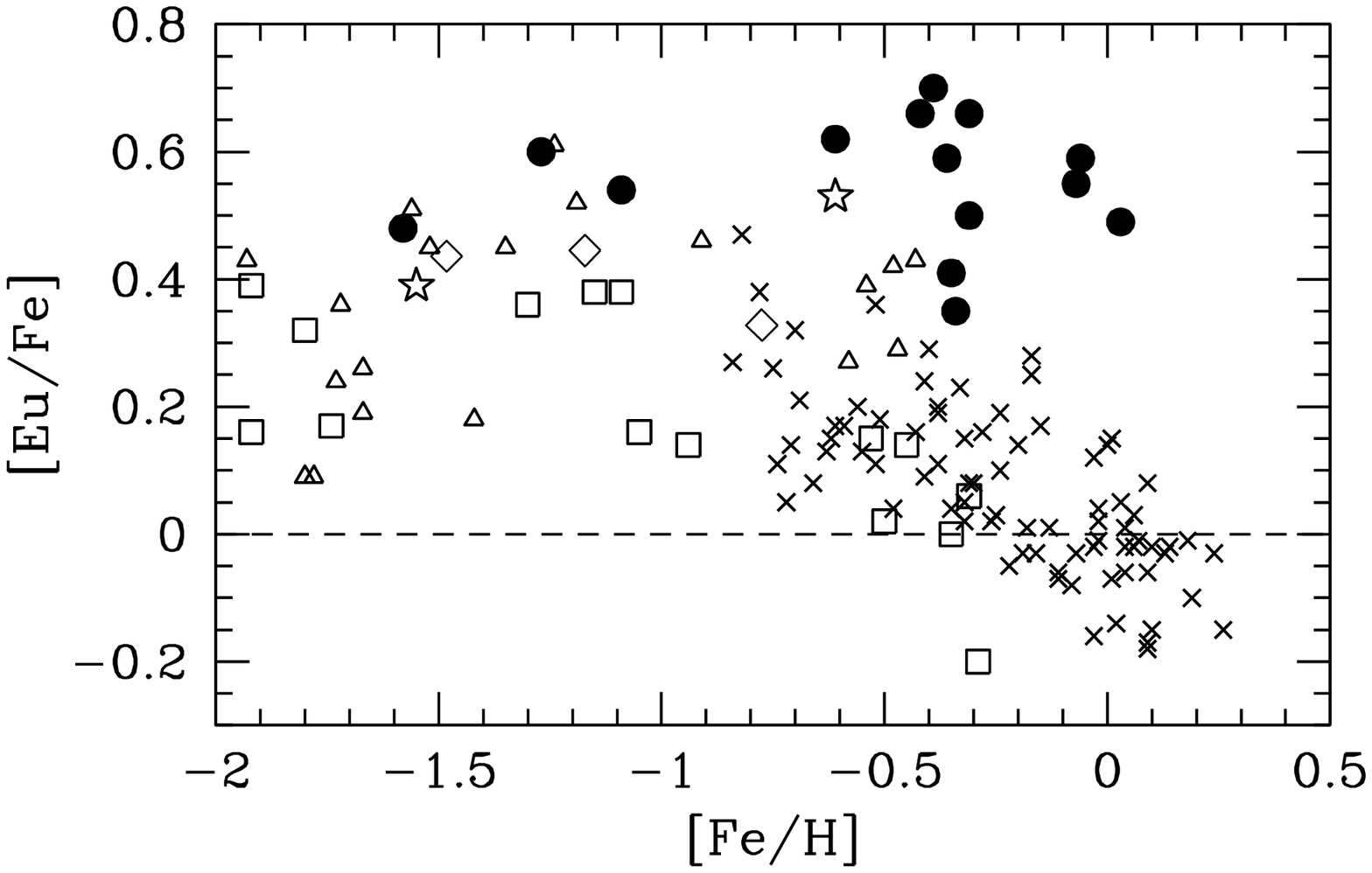}
\caption{ A plot of [Eu/Fe] for the Sgr dSph stars (filled circles), compared
to the solar neighborhood disk and halo: Woolf et al. (1995, crosses),
 Gratton \& Sneden (1994, open squares), Shetrone (1996, clusters: open squares,
field stars: open triangles); other points as in Figure~2.}
\end{figure}

\begin{figure}[!ht]
\plotone{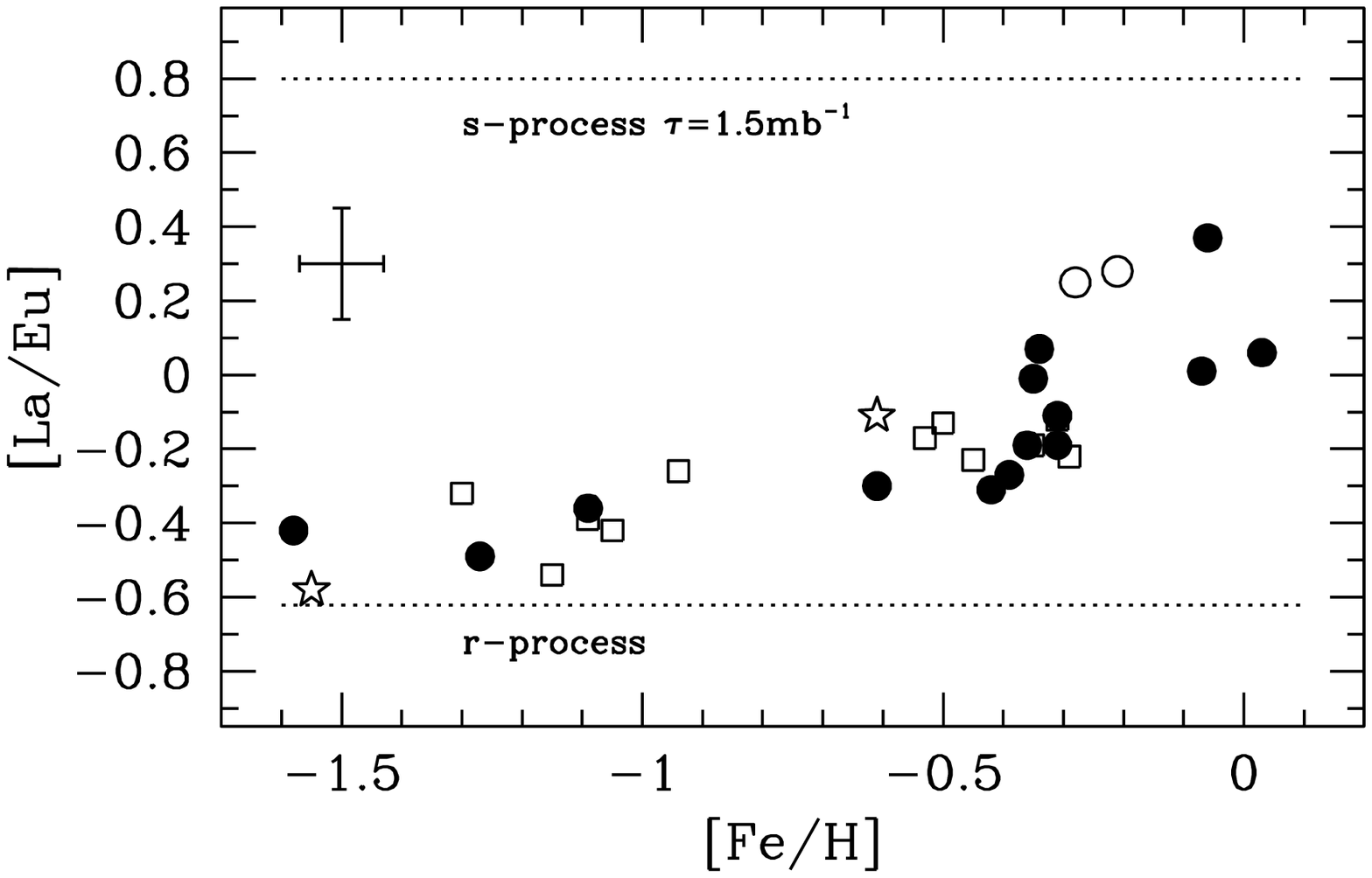}
\caption{ A plot of [La/Eu] versus [Fe/H].  Dotted lines indicate the
solar $r$--process ratio, and a pure $s$--process ratio from Malaney (1987).
Symbols are the same as in Figure~2.
Open squares represent chemical abundances
of Galactic stars from Gratton \& Sneden (1994).}
\end{figure}

%One problem with the previous plots of neutron-capture elements with [Fe/H] is
%that we are not confident about the source of the iron.  

It is interesting to plot [La/Eu] with [La/H], rather than [Fe/H], as the indicator of 
metallicity, as in Figure~8: in this case both axes are related, and it becomes possible 
to make useful
model fits to the data.   In Figure~8 we show three predictions: The solid line
shows the locus of [La/Eu] versus [La/H] when pure s-process, with $\tau$=1.5mb$^{-1}$,
composition is added to the halo composition, at [La/H]=$-$0.8 dex; clearly this does not
fit the observations.  The dashed line shows the locus when pure s-process material
is added to material with the halo [La/Eu] value at [La/H]=$-$0.3 dex; this line
gives a very reasonable fit to the data for [La/H]$\geq$$-$0.3 dex.  The dot-dashed
line shows the locus when 95\% s-process plus 5\% r-process ratio material is added
to the halo composition at [La/H]=$-$0.3 dex.  The 95\% s-process locus also
fits the data reasonably well, but it is clear that a smaller s-process fraction
will give unacceptable fits to the data.  The plots make predictions about the
[La/Eu] ratios for stars of higher [La/H], irrespective of [Fe/H].  The main conclusion
to be drawn from this figure is that a model of chemical evolution in which pure, or
nearly pure, s-process material is added beginning at [La/H]=$-$0.3 dex can explain
those stars with [La/H]$\geq$$-$0.3 dex (corresponding to [Fe/H]$\geq$$-$0.6 dex).

\begin{figure}[!ht]
\plotone{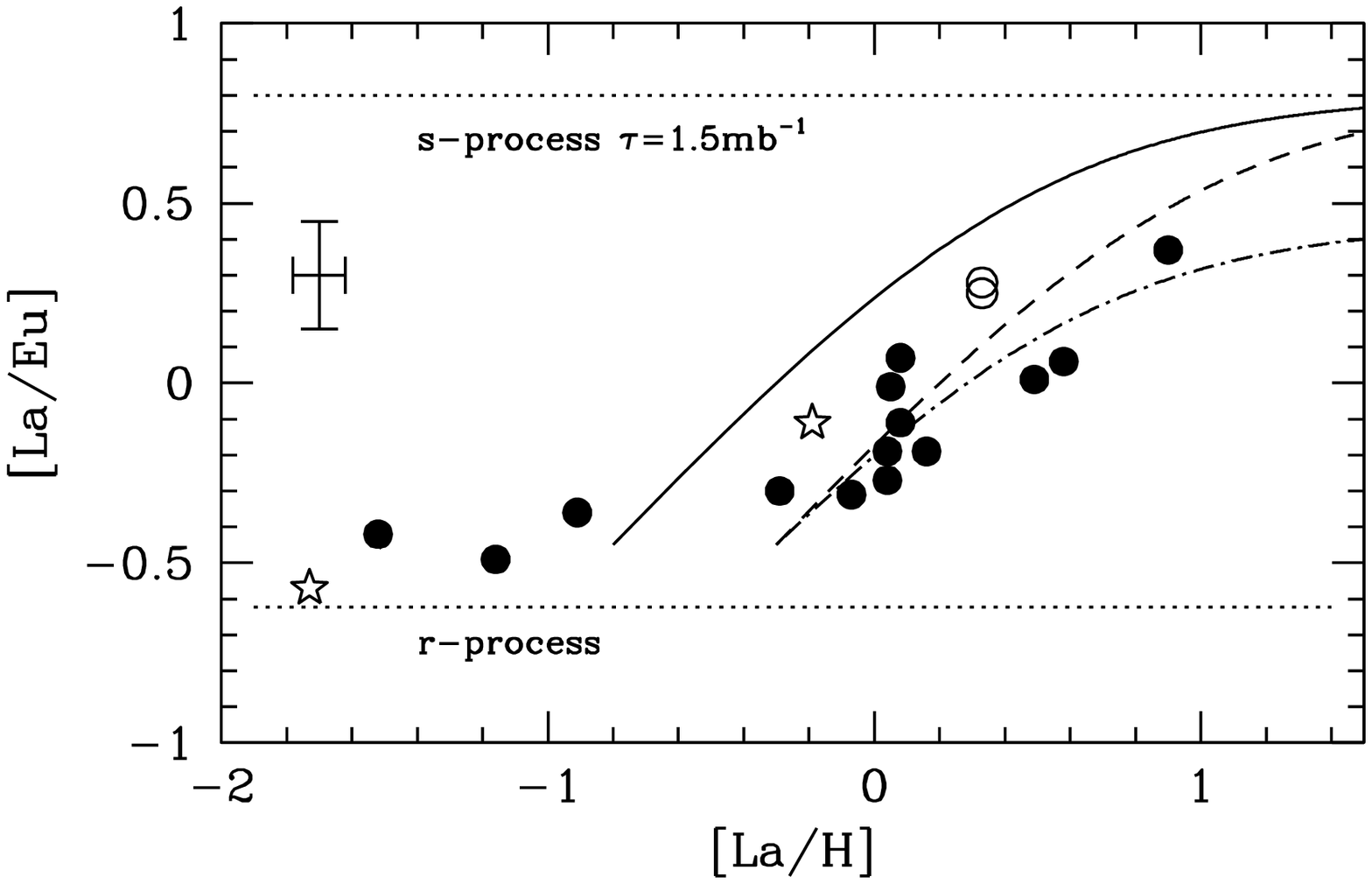}
\caption{ A plot of [La/Eu] versus [La/H].  Dotted lines indicate the
solar $r$--process ratio, and a pure $s$--process ratio from Malaney (1987).
Symbols are the same as in Figure~2.  The solid curve shows the
composition locus when pure $s$-process material is added to halo composition
at [La/H]=$-$0.8; the dashed line shows the locus for pure $s$-process added
to halo composition at [La/H]=$-$0.3, and the dot-dashed line shows the result
of adding 95\% $s$-process and 5\% $r$-process composition to the halo mixture.}
\end{figure}

Figure~9 shows the [La/Y] abundance ratio with [Fe/H].  Clearly [La/Y]
increases between the metal-poor ([Fe/H]$\leq$$-$1) and metal-rich ([Fe/H]$\geq$$-$0.6)
Sgr dSph stars; although the metal-rich stars in our sample appear to show a
roughly constant [La/Y] value, near $+$0.5 dex.  The metal-poor Sgr dSph stars
appear halo-like in their [La/Y] ratios.  We note that Bonifacio et al. (2000) reported
[La/Y] values for their two Sgr dSph stars almost 0.5 dex higher than our sample
at similar [Fe/H].  This difference is due to much lower yttrium abundances for
the two Bonifacio et al. (2000) stars, and might be explained if our Y~II lines
are blended; however, we made extensive efforts using spectrum synthesis of the Kurucz
line list to check that our Y~II lines were, indeed, clean.  We conclude that
either unknown blends have equally affected our three Y~II lines, or there is
some other unknown systematic effect, perhaps due to $gf$ values, which
resulted in the difference between our Y~II abundances and those of Bonifacio et al. (2000).

Figure~9 is especially important because [La/Y] gives a measure of the heavy/light
s-process ratio ([hs/ls]).  This ratio has been shown to be sensitive to the
metallicity in AGB s-process nucleosynthesis (e.g. Busso, Gallino, \& Wasserburg 1999).  At low
metallicity the neutron-to-seed ratio is high in the s-process region, and as a result
the seed nuclei capture many more neutrons than would occur at higher metallicity,
giving a larger production of the heavier neutron-capture nuclei, like La.
In this way one can use the [hs/ls] ratio, in this case [La/Y], to set approximate
constraints on the metallicity of the AGB star responsible for the nucleosynthesis.
Because the [La/Y] ratio is enhanced in the metal-rich Sgr stars, it would
appear that the s-process neutron-capture elements were formed by low-metallicity AGB stars.
This means that the s-process enhancements could not have been made by the stars themselves,
nor could they have been produced by evolved companions, as seen in the solar neighborhood
barium stars.  The AGB s-process nucleosynthesis models of Busso et al. (1999) provide 
a relationship between [Fe/H] and [hs/ls], which we can use to provide a rough estimate
of [Fe/H] of the AGB stars which produced the s-process material present in the metal-rich Sgr 
stars: the bi-valued nature of the relationship suggests that the AGB stars
had [Fe/H]=$-$0.6 or [Fe/H]$\leq$$-$1.4; alternatively one could imagine a range of
AGB [Fe/H] values, heavily weighted to metal-poor stars with [Fe/H]$\leq$$-$0.6 dex.

\begin{figure}[!ht]
\plotone{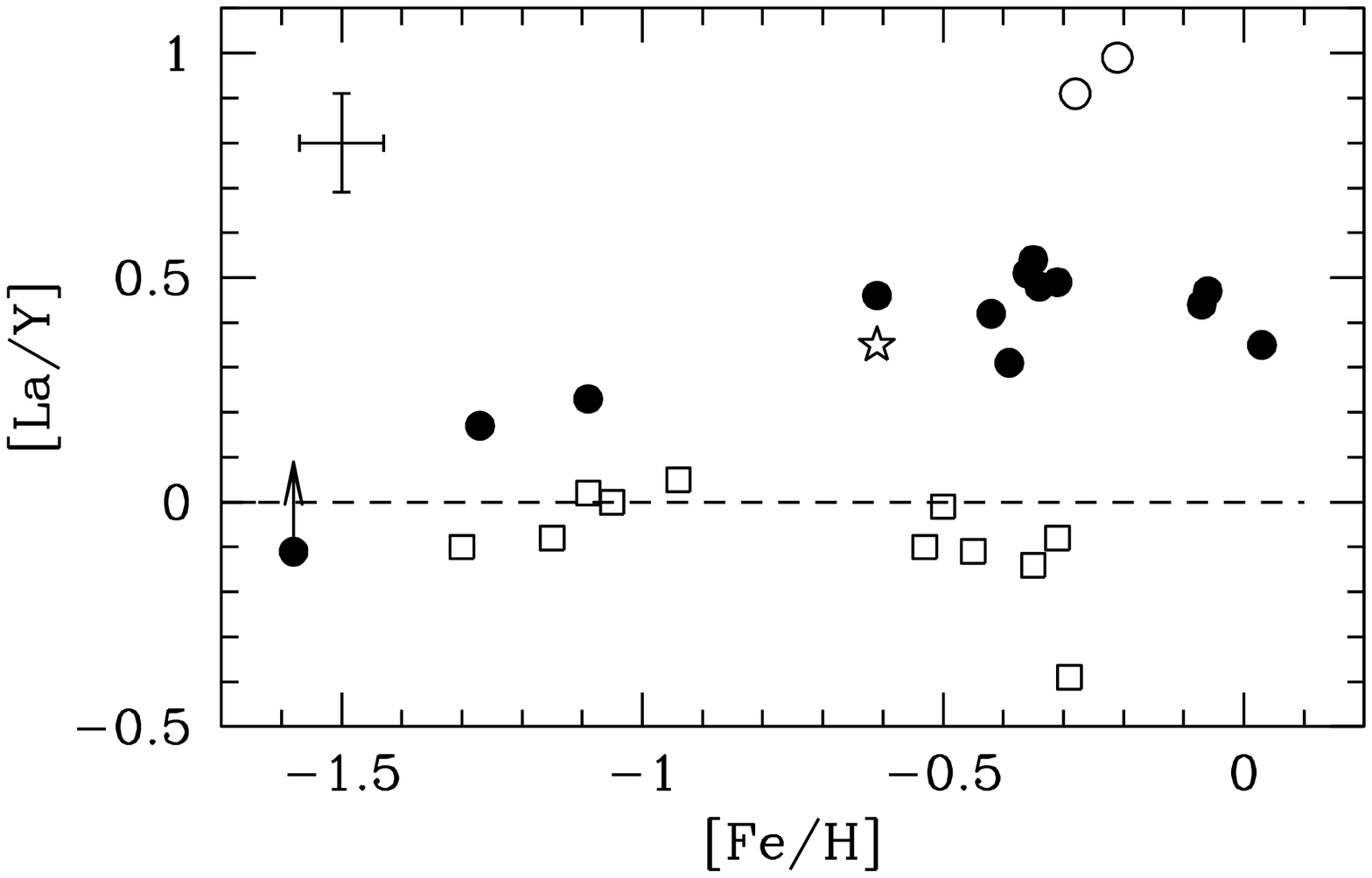}
\caption{A plot of [La/Y] versus [Fe/H].
Symbols are the same as in Figure~2.
Open squares represent chemical abundances
of Galactic stars from Gratton \& Sneden (1994).}
\end{figure}

%\begin{figure}[!ht]
%\plotone{bussohsls.ps}
%\end{figure}

We now turn our attention to the iron-peak elements in Sgr dSph:  Figure~10
shows the trend of [Mn/Fe] in the Sgr dSph and the Galactic bulge, as reported by 
McWilliam et al. (2003).  Great care was taken to try to put the results for both 
onto the same system, and to normalize the results for the solar neighborhood taken
from published studies.  Regarding the solar neighborhood trend for [Mn/Fe]: it has
has long been known (e.g. Wallerstein 1962, Gratton 1989) that [Mn/Fe] declines with [Fe/H],
down to approximately $-$0.4 dex near [Fe/H]=$-$1.  Gratton (1989) proposed that
this trend, being the mirror of the $\alpha$-element enhancements over the
same metallicity range, might be due to an over-production of Mn in type~Ia SN; thus,
at early times, when fewer type~Ia SN had occurred, the [Mn/Fe] ratios were lower.
An alternative explanation for the low [Mn/Fe] ratios in low metallicity stars comes
from the predictions of metallicity-dependent supernova nucleosynthesis yields for
Mn (e.g. Arnett 1971, Woosley \& Weaver 1995); although, these calculations were for type~II
SN nucleosynthesis.

In the Gratton scenario of Mn synthesis low Mn/Fe ratios are associated with enhanced 
$\alpha$/Fe ratios.  Thus, for the Sgr dSph, which as we have shown in Figures~2 and 3,
possess deficient $\alpha$/Fe ratios, we would expect an enhancement in Mn/Fe. This
is opposite from the observed Mn/Fe ratios seen in Figure~10, which are lower than the 
normal trend by $\sim$0.2 dex.   In the Galactic bulge,
where stars have been shown to have $\alpha$/Fe enhancements up to solar [Fe/H]
(e.g. McWilliam \& Rich 1994), Gratton's scenario would predict a decrease in [Mn/Fe];
but as Figure~10 shows, this is not the case: the bulge trend is similar to that of
the solar neighborhood.  

McWilliam et al. (2003) suggested that the sensitivity of 
the Mn yields to neutron excess, or metallicity, might be expected to apply to
type~Ia nucleosynthesis of the iron-peak.  If this is the case, then the low Mn/Fe ratios 
seen in the metal-rich Sgr dSph stars could be explained if the iron-peak material was 
synthesized by low-metallicity type~Ia SN events.  The normal [Mn/Fe] of the bulge stars 
would also be explained by a metallicity-dependent yield, despite the $\alpha$-element
enhancements.  

Timmes, Brown, \& Truran (2003) argue that the $^{56}$Ni yield
from SN~Ia is modulated by the progenitor metallicity, in the 0.2 to 0.8 M$_{\odot}$
mass shells, as a result of a rapid burn to nuclear statistical equilibrium (NSE), and
they predict a subsequent metallicity-dependent variation of the peak luminosity.
The same argument indicates that low metallicity (low neutron excess) NSE
would also reduce the yield of the neutron-rich isotopes, such as Mn.

We arrive at a picture where both the iron-peak elements and the neutron-capture
elements in metal-rich Sgr stars have been made by a much more metal-poor population.
McWilliam et al. (2003) predicted that if this is the case, then the element ratios
of other metallicity-dependent species should be similarly affected; in particular,
the [Cu/Fe] trend in the Sgr dSph should be deficient relative to the solar neighborhood,
since theoretical predictions (e.g. Woosley \& Weaver 1995) and observations
(e.g. Sneden \& Crocker 1988, Mishenina et al. 2002, Simmerer et al 2003) suggest a
strongly metallicity-dependent Cu yield.

\begin{figure}[!ht]
\plotone{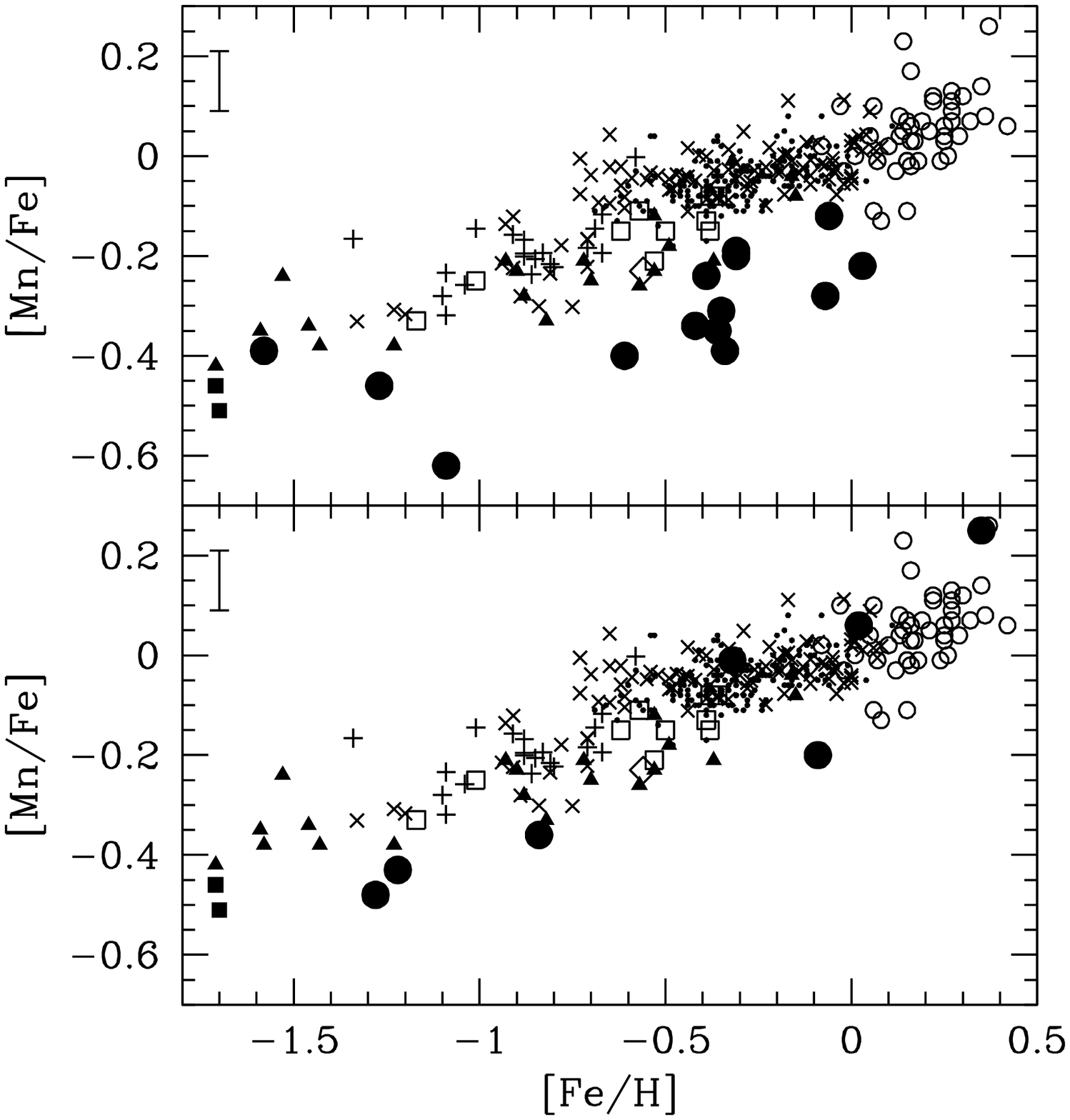}
%\plotone{mnfe_sgr.eps}
%\plotone{mnfe_bulge.eps}
\caption{ {\em Upper:} A plot showing [Mn/Fe] with [Fe/H] in Sgr dSph compared to the
solar neighborhood. \quad\quad
{\em Lower:} The same plot, but for [Mn/Fe] in a sample of Galactic bulge stars.  
See McWilliam et al. (2003) for details.}
%\plottwo{mnfe_bulge.eps}{mnfe_bulge.eps}
%\plotfiddle{mnfe_bulge.eps}{10.0cm}{0}{40}{40}{0}{0}
%\plotfiddle{mnfe_sgr.eps}{10.0cm}{0}{40}{40}{-260}{300}
\end{figure}

In Figure~11 we show the observed [Cu/Fe] versus [Fe/H] for our 14 Sgr dSph red giant
stars, as reported by McWilliam \& Smecker-Hane (2004).  The abundances are
based on the Cu I line at 5782\AA , include appropriate $hfs$ effects, and are
normalized to the result obtained from the solar line.  The low metallicity Sgr dSph
stars have [Cu/Fe] values typical of the Galactic halo, whilst [Cu/Fe] is deficient by 
up to 0.5 dex in the metal-rich Sgr stars, with a hint of an increase in [Cu/Fe] for 
the highest metallicity stars.
Note that the abundance for $\alpha$~Boo suggests that, if anything, systematic errors
may require us to lower our [Cu/Fe] ratios for the Sgr dSph.

The low [Cu/Fe] ratios seen in Figure~11 support the conclusions from the [Mn/Fe]
abundance ratios; namely, that the yields of both these elements are
metallicity-dependent, and that in the metal-rich Sgr dSph the iron-peak abundances 
are dominated by the yields from of a much lower metallicity population.

\begin{figure}[!ht]
\plotone{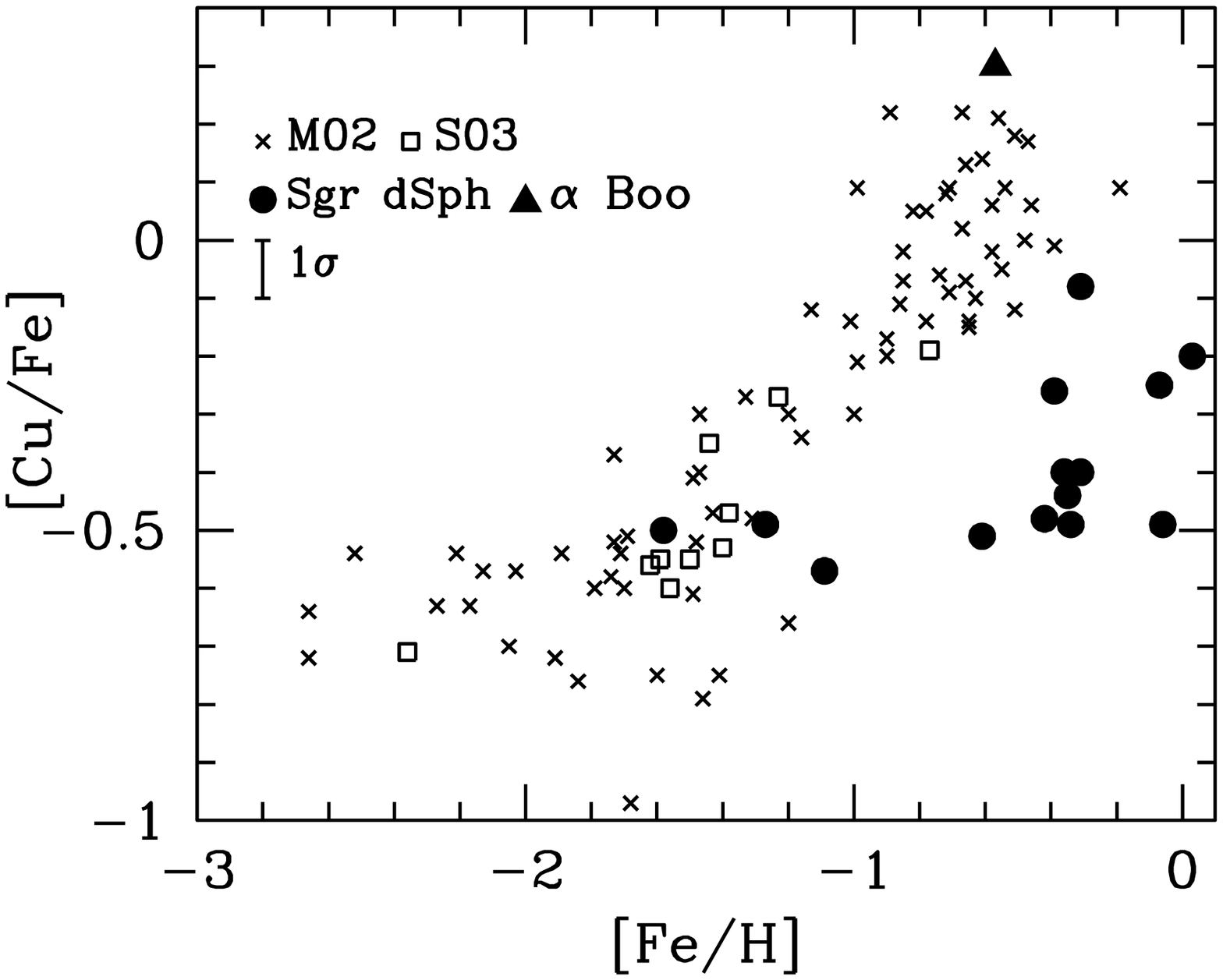}
\caption {[Cu/Fe] values for 14 Sgr dSph stars (filled circles),
compared to the trend for Galactic disk and halo stars, as measured by
Mishenina et al. (2002; M02 crosses) and ten globular clusters 
(Simmerer et al.  2003; S03 open squares).  The filled triangle is for Arcturus, 
as measured by McWilliam \& Smecker-Hane (2004). }
\end{figure}

\section{Discussion}

We would now like to produce a scenario for the evolution of the Sgr dSph
to explain all the observed abundance trends, outlined in the
previous section.

The neutron-capture element abundances suggest that metal-poor AGB stars played a 
significant role in the synthesis of the heavy elements.  Since AGB stars that produce
s-process elements are usually of low mass (typically 1.3 to 3M$_{\odot}$) the progenitors 
are probably long-lived, and we conclude that an old, metal-poor, population dominates
the abundances of s-process elements in the younger metal-rich stars.  

The low [$\alpha$/Fe] ratios, long time scales inferred by the presence of AGB
s-process material, and the large spread in ages of Sgr dSph stars, inferred from
color-magnitude diagrams (e.g. Layden \& Sarajedini 2000), suggest that type~Ia
SN contributed significantly to the composition of the younger, metal-rich
Sgr population.

The low abundances of Mn and Cu suggest that the iron-peak elements of the metal-rich
population were synthesized by low-metallicity supernovae, which must have been dominated
by type~Ia, since type~II SN, with progenitors formed from the metal-rich young gas, would 
have produced  normal Mn and Cu abundances.

The importance of metal-poor type~Ia, from old progenitors, in the nucleosynthesis
of the metal-rich Sgr dSph material is supported by the observed deficiencies
of Na and Al, since these two elements are expected to be under-produced in type~Ia
SN, and under-produced at low metallicity.

An important question to answer is: How did synthesis products from the old, metal-poor,
population come to dominate the composition of the metal-rich stars?
For the injection of metals from an old, metal-poor, population to
exceed the injection from young stars it is necessary to have a relatively low
star formation rate at late times.  This could occur in at least two ways:

First, through a significant loss of gas from the galaxy over long time scales.
In this situation the initial galactic formation event produced
a large number of low mass, low metallicity, stars.  Subsequent mass loss, 
over several Gyr, would be expected of a low-mass dwarf galaxy like the Sgr dSph, 
and would leave a small pool of gas to form the, younger, metal-rich population.  
Over this long period of time the old, low-metallicity, population would inject a
large amount of gas, both from the envelopes of expired AGB stars, and through
ejecta from type~Ia SN.  This material could overwhelm the material injected
from the younger generations; thus, the new generations of stars would reflect
the synthesis products of the old, metal-poor, stars, as observed.

A second possible scenario is that the Sgr dSph experienced chemical
enrichment dominated by a powerful burst of star formation, followed by a quiescent
period of several Gyrs.  Thus, after the major burst of star formation the injection of
newly synthesized material from type~Ia SN and AGB stars had time to occur, and could
have overwhelmed the new material from young stars formed in the quiescent period.
This second scenario would likely leave a reservoir of gas, which
would have to be lost from the galaxy at a later time.

%A second possible scenario is that the Sgr dSph experienced chemical
%enrichment dominated by a small number of bursts of star formation, separated by
%several Gyrs.  Thus, during the time between bursts the injection of newly synthesized
%material from type~Ia SN and AGB stars had time to occur, and could have overwhelmed
%the new material from young stars formed in a low level of on-going star formation
%between bursts.

It is likely that a combination of both these scenarios operated in the Sgr dSph.
To estimate the relative importance of these two scenarios it would be useful to
have an unbiased metallicity function of the entire system.
If the mass-loss scenario dominated we might expect the metallicity
function to be biased to low-metallicity stars, while a discrete star-formation burst
might lead to a metallicity function dominated 
by a narrow peak; but detailed models of the metallicity evolution would be
required to distinguish between the two scenarios.

Observations of the metallicity function from Bellazzini et al. (1999) found 
8 to 30\% of Sgr dSph stars, depending on location, with [Fe/H]$\geq$$-$0.7 dex.
On the other hand, Cole (2001) suggested that the Sgr dSph metallicity is peaked at
[Fe/H]$\sim$$-$0.5$\pm$0.2 dex, based on 2MASS photometry.

Given that extensive mass-loss and, perhaps, star formation bursts are expected for
low mass galaxies it should be the case that other, low-mass, galaxies have experienced
a chemical evolution history similar to Sgr dSph.  Thus, we might expect the
chemical abundance patterns found in the Sgr dSph to be expressed in other low-mass
systems.  Observational evidence for this comes from two systems: the Galactic globular
cluster $\omega$~Cen, and the Fornax dwarf galaxy.

Abundance analyses of stars in $\omega$~Cen (e.g. Lloyd Evans 1983, Smith et al. 2000)
show that the more metal-rich stars in the cluster show [La/H] enhancements 
of $\sim$1 dex, and [La/Y] enhanced by $\sim$0.3--0.4 dex; Cunha et al. (2002)
found deficient [Cu/Fe] ratios in the metal-rich $\omega$~Cen stars by as much as 0.5 dex,
relative to the solar neighborhood trend.  These results are
remarkably similar to our findings for the Sgr dSph, except that $\omega$~Cen has
a lower average [Fe/H], near $-$1.8 dex.  These abundance similarities, as well as
the spread in [Fe/H] suggest that the two systems share similar chemical evolution
histories.  However, $\omega$~Cen shows the normal halo enhancement of the alpha
elements, and enhancements of Na and Al, rather than the deficiencies we find in
Sgr dSph; thus, there must have been some differences in their evolutionary histories.
This supports the idea that $\omega$~Cen is the nucleus of an accreted dwarf galaxy 
(e.g. Gnedin et al. 2002).

Three stars in the Fornax dwarf spheroidal galaxy were analyzed by
Shetrone et al. (2003); two were metal-poor, and possessed halo-like neutron-capture
abundances, but the third star, with [Fe/H]$\sim$$-$0.7 showed large s-process enhancements,
$\sim$1 dex.  Shetrone et al. (2003) concluded that the most likely explanation for 
this star is that it underwent mass transfer in a binary system with an evolved AGB star;
although with such a small sample they could not rule out the possibility that the
most metal-rich stars in Fornax have large s-process enrichment from AGB stars, in
comparison to the number of r-process SN events.
Contrary to Shetrone et al. (2003) our opinion is that the similarity of the s-process
enhancements in metal-rich Sgr dSph and Fornax stars strongly suggest that Fornax 
evolved similar to Sgr dSph.
%experienced a similar evolution history to the Sgr dSph.

AMcW gratefully acknowledges support from NSF grants AST-96-18623 and AST-00-98612.  TSH
acknowledges support from NSF grants AST-96-19460 and AST-00-70895, and a AAS Small Research Grant.

%\begin{figure}[!ht]
%\plotone{vvs_ocen_la.ps}
%\end{figure}

%\begin{figure}[!ht]
%\plotone{vvs_ocen_y2.ps}
%\end{figure}

%\begin{figure}[!ht]
%\plotone{cu_ocen.ps}
%\end{figure}

{}


\begin{thebibliography}{}

\bibitem[Allende Prieto et al. (2004)]{}
Allende Prieto, C., Barklem, P.S., Lambert, D.L., \& Cunha, K. 2004, \aap, 420, 183


\bibitem[Allende Prieto et al. (2001)]{}
Allende Prieto, C., Lambert, D.L., \& Asplund, M. 2001, \apj, 556, L63

\bibitem[Arnett (1971)]{}
Arnett, W.D. 1971, \apj, 166, 153

\bibitem[Arnett \& Thielemann (1985)]{}
Arnett, W.D., \& Thielemann, F.-K. 1985, \apj, 295, 589

\bibitem[Bellazzini et al. (1999)]{}
Bellazzini, M., Ferraro, F.R., \& Buonanno, R. 1999, \mnras, 307, 619

\bibitem[Bonifacio et al. (2000)]{}
Bonifacio, P., Hill, V., Molaro, P., Pasquini, L., Di Marcantonio, P., \& Santin, P. 2000,
       \aap, 359, 663

\bibitem[Brown et al. (1999)]{}
Brown, J.A., Wallerstein, G., \& Gonzalez, G. 1999, \aj, 118, 1245

\bibitem[Busso et al (1999)]{}
Busso, M., Gallino, R., \& Wasserburg, G.J. 1999, \araa, 37, 239

\bibitem[Cole (2001)]{}
Cole, A.A. 2001, \apj, 559, L17

\bibitem[Cunha et al. (2002)]{}
Cunha, K., Smith, V.V., Suntzeff, N.B., Norris, J.E., Da Costa, G.S., \& Plez, B. 2002, 
       \apj, 124, 379

\bibitem[Edvardsson et al. (1993)]{}
Edvardsson, B., Andersen, J., Gustafsson, B., Lambert, D.L., Nissen, P.E., \& Tomkin, J. 1993, 
      \aap, 275, 101

\bibitem[Gnedin et al. (2002)]{}
Gnedin, O.Y., Zhao, G., Pringle, J.E., Fall, M.S., Livio, M., \& Meylan, G.
     2002, ApJ, 568, L23


\bibitem[Gratton (1989)]{}
Gratton, R.G. 1989, \aap, 208, 171

\bibitem[Gratton \& Sneden (1988)]{}
Gratton, R.G., \& Sneden, C. 1988, \aap, 204, 193

\bibitem[Gratton \& Sneden (1994)]{}
Gratton, R.G., \& Sneden, C. 1994, \aap, 287, 927


%\bibitem[Hinkle et al. (2000)]{}
%Hinkle, K., Wallace, L., Valenti, J., \& Harmer, D. 2000,  ``Visible and Near
%   Infrared Atlas of the Arcturus Spectrum 3727-9300 \AA'' eds. Kenneth Hinkle, Lloyd
%   Wallace, Jeff Valenti, and Dianne Harmer. (San Francisco: ASP) ISBN:
%   1-58381-037-4, 2000.


\bibitem[Kraft et al. (1997)]{}
Kraft, R.P., Sneden, C., Smith, G.H., Shetrone, M.D., Langer, G.E., \& Pilachowski, C.A. 1997, 
        \aj, 113, 279

\bibitem[Kurucz (1993)]{}
Kurucz, R.L. 1993, private communication

\bibitem[Layden \& Sarajedini 2000]{}
Layden, A.C., \& Sarajedini, A. 2000, \aj, 119, 1760

\bibitem[Lloyd Evans (1983)]{}
Lloyd Evans, T. 1983, \mnras, 204, 975

\bibitem[Malaney (1987)]{}
Malaney, R.A. 1987, \apj, 321, 832

\bibitem[Matteucci \& Brocato (1990)]{}
Matteucci, F., \& Brocato, E. 1990, \apj, 365, 539

\bibitem[McWilliam (1997)]{}
McWilliam, A. 1997, \araa, 35, 503

\bibitem[McWilliam et al. (1995)]{}
McWilliam, A., Preston, G.W., Sneden, C., \& Shectman, S. 1995, \aj, 109, 2736

\bibitem[McWilliam \& Rich (1994)]{}
McWilliam, A., \& Rich, R.M. 1994, \apjs, 91, 749

\bibitem[McWilliam et al. (2003)]{}
McWilliam, A., Rich, R.M., \& Smecker-Hane, T.A. 2003, \apj, 592, L145

\bibitem[McWilliam \& Smecker-Hane (2004)]{}
McWilliam, A., \& Smecker-Hane, T.A. 2004, \apj, {\em submitted}

\bibitem[Mishenina et al. (2002)]{}
Mishenina, T.V., Kovtyukh, V.V., Soubiran, C., Travaglio, C., \& Busso, M. 2002, 
        \aap, 396, 189

\bibitem[Simmerer et al. (2003)]{}
Simmerer, J., Sneden, C., Ivans, I.I., Kraft, R.P., Shetrone, M.D., \& Smith, V.V. 2003,
       \aj, 125, 2018

\bibitem[Shetrone (1996)]{}
Shetrone, M.D., 1996, \aj, 112, 1517

\bibitem[Shetrone et al. (2003)]{}
Shetrone, M., Venn, K.A., Tolstoy, E., Primas, F., Hill, V., \& Kaufer, A. 2003, \aj, 125, 684

\bibitem[Smecker-Hane \& McWilliam (2002)]{}
Smecker-Hane. T.A., \& McWilliam, A. 2002, astro-ph/0205411  (SM02)

%\bibitem[Smecker-Hane \& McWilliam (2005)]{}
%Smecker-Hane. T.A., \& McWilliam, A. 2005, in progress

\bibitem[Smith et al. (2000)]{}
Smith, V.V., Suntzeff, N.B., Cunha, K., Gallino, R., Busso, M., Lambert, D.L., \& 
     Straniero, O.  2000, \aj, 119, 1239

\bibitem[Sneden \& Crocker (1988)]{}
Sneden, C., \& Crocker, D.A. 1988, \apj, 335, 406

\bibitem[Sneden et al. (2004)]{}
Sneden, C., Kraft, R.P., Guhathakurta, P., Peterson, R.C., \& Fulbright, J.P. 2004, 
      \aj, 127, 2162

\bibitem[Tautvaisiene et al. (2004)]{}
Tautvaisiene, G., Wallerstein, G., Geisler, D., Gonzalez, G., \& Charbonnel, C. 2004, 
       \aj, 127, 373

\bibitem[Timmes et al. (2003)]{}
Timmes, F.X., Brown, E.F., \& Truran, J.W. 2003, \apj, 590, L83

\bibitem[Tinsley (1979)]{}
Tinsley, B.M. 1979, \apj, 229, 1046

\bibitem[Vogt et al. (1994)]{}
Vogt, S.S. et al. 1994, S.P.I.E., 2198, 362

\bibitem[Wallerstein (1962)]{}
Wallerstein, G. 1962, \apjs, 6, 407

\bibitem[Wheeler et al. (1989)]{}
Wheeler, J.C., Sneden, C., \& Truran, J.W. 1989, \araa, 27, 279

\bibitem[Woolf, V. et al. 1995]{}
Woolf, V., Tomkin, J., \& Lambert, D.L. 1995, \apj, 453, 660 

\bibitem[Woosley \& Weaver (1995)]{}
Woosley, S.E., \& Weaver, T.A. 1995, \apjs, 101, 181

\end{thebibliography}
\end{document}